\documentclass[aip,apl,amsmath,amssymb,reprint,superscriptaddress]{revtex4-1}
\usepackage{graphicx}

\begin{document}

\title{Polar POLICRYPS Diffractive Structures Generate Cylindrical Vector Beams}

\author{Domenico Alj}%
\affiliation{Department of Physics \& CNR-NANOTEC University of Calabria, I-87036 Rende (CS), Italy}%
\author{Sathyanarayana Paladugu}%
\affiliation{Soft Matter Lab, Department of Physics, Bilkent University, Ankara 06800, Turkey.}%
\author{Giovanni Volpe}%
\affiliation{Soft Matter Lab, Department of Physics, Bilkent University, Ankara 06800, Turkey.}
\affiliation{UNAM-National Nanotechnology Research Center, Bilkent University, Ankara 06800, Turkey.}
\author{Roberto Caputo}%
 \email{roberto.caputo@fis.unical.it}
\affiliation{Department of Physics \& CNR-NANOTEC University of Calabria, I-87036 Rende (CS), Italy}
\author{Cesare Umeton}
\affiliation{Department of Physics \& CNR-NANOTEC University of Calabria, I-87036 Rende (CS), Italy}

\date{\today}

\begin{abstract}
Local shaping of the polarization state of a light beam is appealing for a number of applications. This can be achieved by employing devices containing birefringent materials. In this article, we present one such device that permits one to convert a uniformly circularly polarized beam into a cylindrical vector beam (CVB). This device has been fabricated by exploiting the POLICRYPS photocuring technique. It is a liquid-crystal-based optical diffraction grating featuring polar symmetry of the director alignment. We have characterized the resulting CVB profile and polarization for the cases of left and right circularly polarized incoming beams.
\end{abstract}


\maketitle

The spatial modulation of the optical axis orientation of a birefringent material can be employed to build novel devices capable of locally modify the state of polarization (SOP) of a light beam. For example, a light beam entering such a device with uniform polarization can exit with a tunable SOP that is spatially variable over the beam area. This holds the promise of exploiting novel effects and phenomena that can expand the functionalities and enhance the capabilities of optical systems. In fact, this is the core of Tabiryan's idea for optics of the $4^{\rm th}$ generation~\cite{Tabirian}, following optics based on simple shape ($1^{\rm st}$ generation), refractive index ($2^{\rm nd}$ generation) and effective birefringence ($3^{\rm rd}$ generation). As this kind of devices shape the SOP of a beam, we will call them \emph{polarization shapers}, or \emph{polshapes}. In this article, we report on the design and realization of a specific polshape that is able to convert a circularly polarized beam into a cylindrical vector beam (CVB). 

\begin{figure}[!h]
\centering
\includegraphics[width=\linewidth]{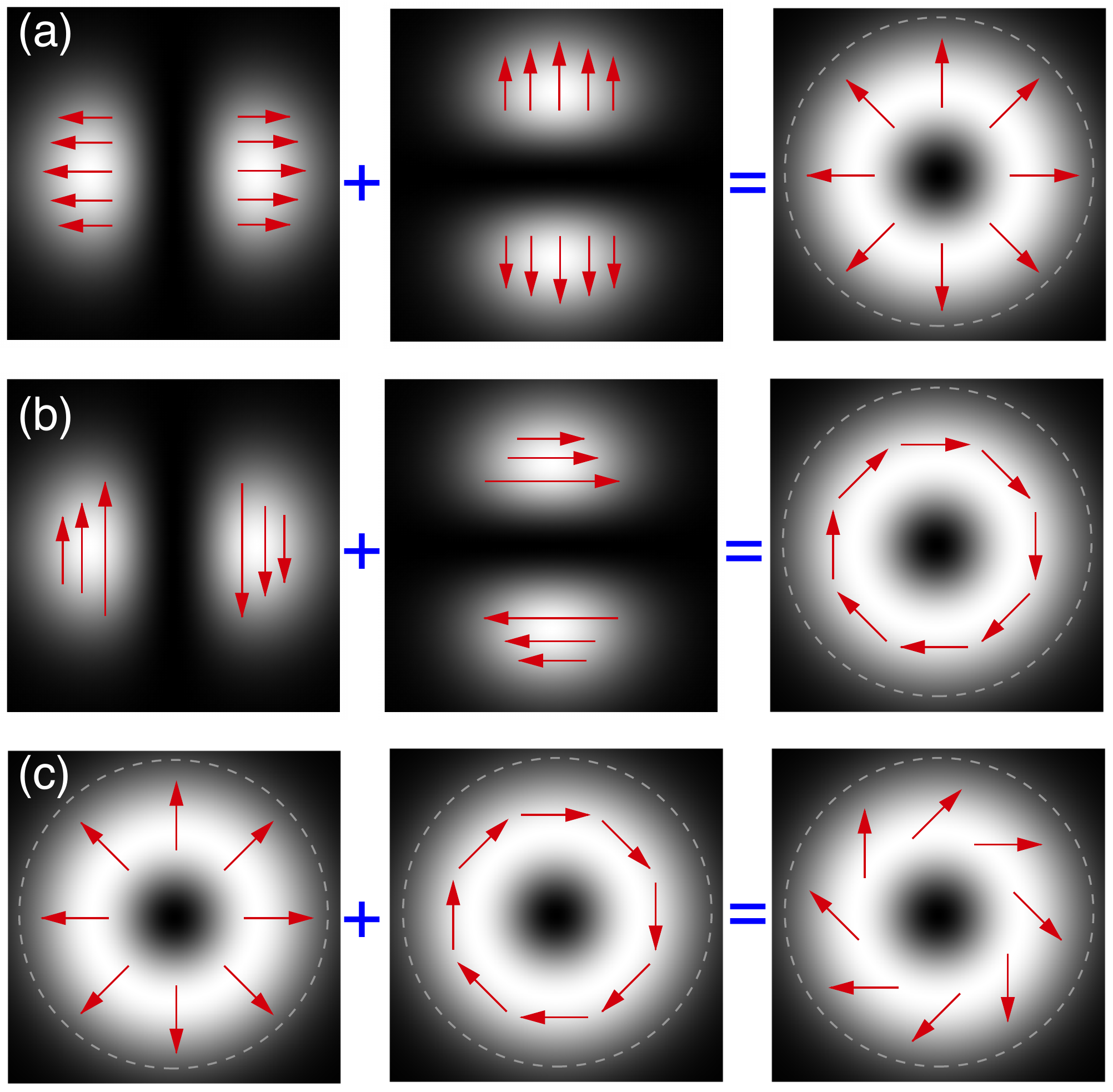}
\caption{(Color online) Cylindrical vector beams (CVBs). (a) Radially and (b) azimuthally polarized $1^{\rm st}$-order cylindrical vector beams can be considered as the linear superposition of orthogonally polarized $1^{\rm st}$-order Hermite-Gaussiam beams. (c) A hybrid CVB can be obtained as the superposition of a radial CVB and an azimuthal CVB. The arrows represent the local direction of the polarization vector.}
\label{fig:fig1}
\end{figure}

CVBs are vector solutions of Maxwell's equations that obey axial symmetry in amplitude, phase and polarization~\cite{Hall,Zhan}. As shown in Fig.~\ref{fig:fig1}, the simplest $1^{\rm st}$-order CVBs can be expressed as the superposition of two ortogonally polarized $1^{\rm st}$-order Hermite-Gaussian (HG) beams~\cite{Zhan}. A radially polarized CVB is obtained as (Fig.~\ref{fig:fig1}a)
\begin{equation}\label{eqn:eqn5}
\bold{E}^{CV}_r(x,y,z)={\rm HG}_{10}(x,y,z)\hat{x}+{\rm HG}_{01}(x,y,z)\hat{y} \; ,
\end{equation}
and an azimuthally polarized CVB as (Fig.~\ref{fig:fig1}b)
\begin{equation}\label{eqn:eqn6}
\bold{E}^{CV}_a(x,y,z)={\rm HG}_{10}(x,y,z)\hat{x}-{\rm HG}_{01}(x,y,z)\hat{y} \; .
\end{equation}
Also other $1^{\rm st}$-order CVB exist, e.g., a superposition of the radial and azimuthal CVBs leads to the hybrid CVB shown in Fig.~\ref{fig:fig1}c. The interest in CVBs arose mainly because of their unique focusing properties due to their polarization symmetry~\cite{Zhan1,Zhan2}. A radially polarized CVB can be focussed into a tighter spot~\cite{Quabis,Dorn} when compared to a Gaussian beam, while the focus of an azimuthally polarized CVB has a donut shape. This has been exploited, e.g., in optical trapping~\cite{Jones}, where radial CVBs have been used to trap metallic particles~\cite{Zhan3,Zhan4} and azimuthal CVBs to trap particles with dielectric constant lower than that of the ambient medium~\cite{Zhan2}. Switching between radial, azimuthal and hybrid polarization can be done using two half-wave plates~\cite{Zhan5,Volpe}. Various alternative methods have been proposed to produce CVBs including: 
a double-interferometer configuration to convert a linearly polarized laser beam into a radially polarized one~\cite{Tidwell}; 
a radial analyzer consisting of a birefringent lens~\cite{Zhan5}; 
a surface-emitting semiconductor laser~\cite{Erdogan}; a space-varying liquid crystal cell~\cite{Stalder}; 
the summation inside a laser resonator of two orthogonally polarized ${\rm TEM}_{01}$ modes~\cite{Oron}; 
the excitation of a few-modes optical fiber with an offset linearly polarized Gaussian beam ~\cite{Grosjean};
a space-variant dielectric subwavelength grating~\cite{Bomzon}; 
a few-modes fiber excited by a Laguerre-Gaussian beam~\cite{Volpe};
a conical Brewster prism~\cite{Kozawa}; 
and a Sagnac interferometer~\cite{Niziev}.

\begin{figure}[!h]
\centering
\includegraphics[width=\linewidth]{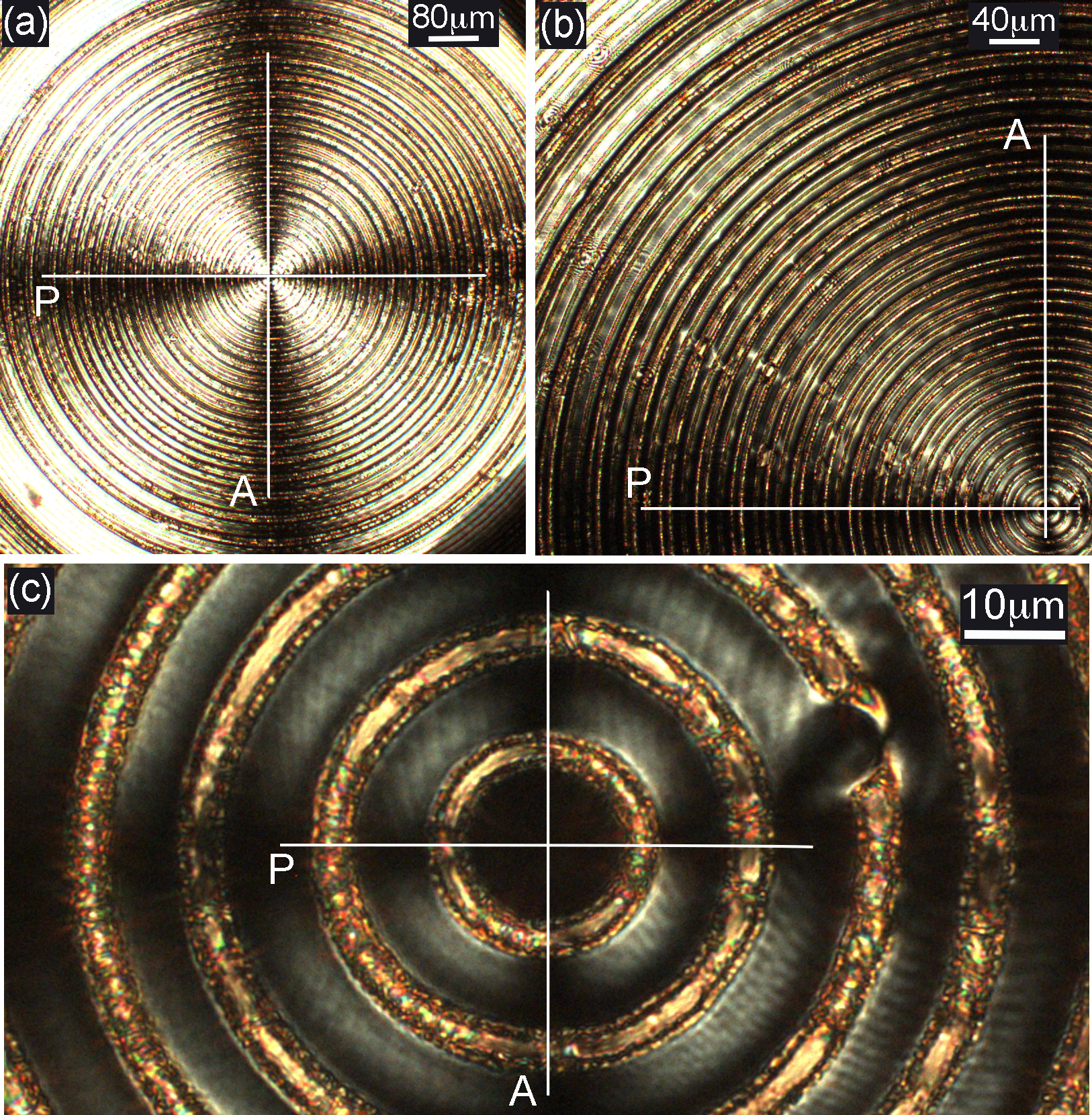}
\caption{(Color online) Micrographs of a POLICRYPS polshape with radial symmetry. All the images are acquired with a polarized optical microscope between crossed polarizers and represent the same sample but at different magnifications. The yellow/light grey rings are the polymeric circles and the dark grey rings are the aligned LC regions.}
\label{fig:fig2}
\end{figure}

We have realized the CVB polshape employed in this work by the one-step POLICRYPS photocuring technique described in detail in Refs.~\citenum{Caputo,Infusino,Sio}. The fabricated sample is a polar diffractive structure that contains a polymeric support structure and a birefringent material, i.e., liquid crystals (LCs), whose optical axis is radially oriented. This radial symmetry matches the cylindrical symmetry of CVBs (see Fig.~\ref{fig:fig1}) and, thus, led us to expect that such polshape could generate a CVB. The exact fabrication procedure is reported in Ref.~\citenum{Alj}. In brief, we have followed the ensuing steps:
\begin{enumerate}
\item A curing laser-light pattern with the shape of concentric rings induces phase separation between the polymer and LC molecules. The resulting formation of polymeric rings replicates the light pattern and drive the diffusion of LC molecules in the regions comprised between rings.
\item The photocuring process takes place at high temperature (e.g., above the nematic-isotropic transition temperature of the LCs) as in a typical POLICRYPS fabrication procedure~\cite{Marino}. This ensures that the LCs do not separate in droplets but remains homogeneous between the polymeric rings and their directors orient  perpendicularly to the rings in a radially symmetric configuration.  
\end{enumerate}
Some micrographs of the obtained structure are reported in Fig.~\ref{fig:fig2}. All images represent the same sample, but at different magnifications, as observed at the polarized optical microscope between crossed polarizers. The  Maltese cross confirms the expected radial alignment of the LC molecules between the polymeric rings of the polshape~\cite{Sio1,Caputo1}. Interestingly, the high magnification detail of the center (Fig.~\ref{fig:fig2}c) reveals the contrast between the polymeric circles (yellow/light grey rings) and the aligned (dark grey) LC regions.

\begin{figure}[!h]
\centering
\includegraphics[width=\linewidth]{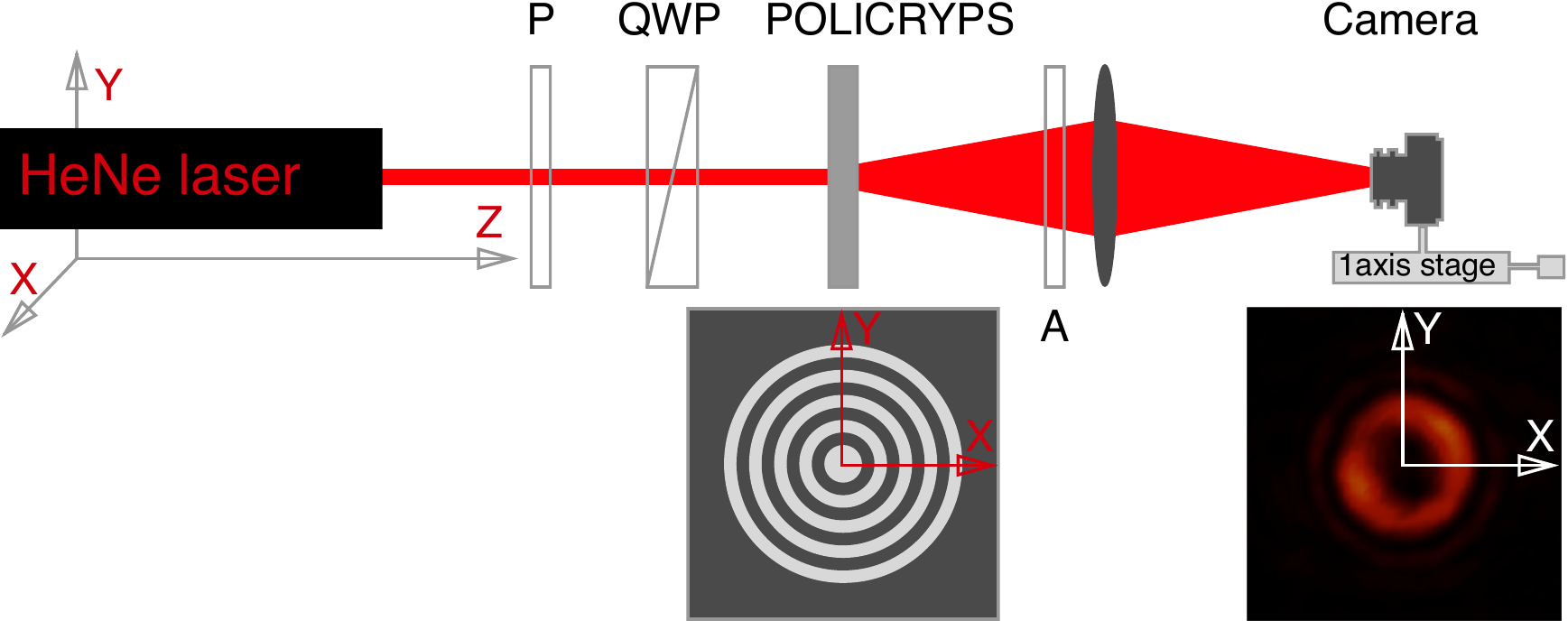}
\caption{(Color online) Setup employed to generate CVBs using the POLICRYPS polshape. It includes a He-Ne laser source ($633\,{\rm nm}$), a linear polarizer (P), a quarter-wave plate (QWP), a POLICRYPS polshape, a lens, another (removable) polarizer (A) and a camera monted on a $z$-axis translation stage. 
}
\label{fig:fig3}
\end{figure}

\begin{figure*}[t!]
\centering
\includegraphics[width=\linewidth]{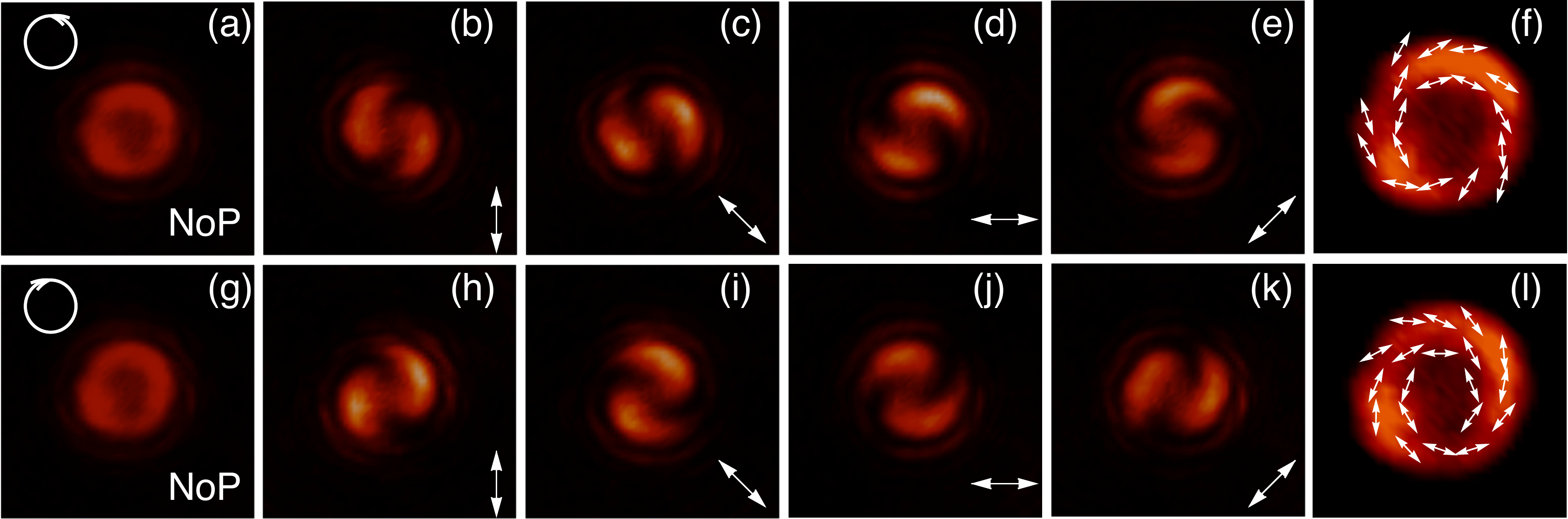}
\caption{(Color online) CVBs generated using a POLICRYPS polshape illuminated with (a)-(f) a LCP Gaussian beam and (g)-(l) a RCP Gaussian beam. (a) and (g) show the resulting beams emerging from the polshape, focused by the lens and imaged by the camera (see setup in Fig.~\ref{fig:fig3}). (b)-(e) and (h)-(k) show the beams after an analyzer (i.e., a linear polarizer) oriented along the direction of the arrow. (f) and (l) depict the polarization states measured using Stokes parameters.}
\label{fig:fig4}
\end{figure*}

The setup used to generate CVBs using our POLICRYPS polshape is shown in Fig.~\ref{fig:fig3}. The laser source is a He-Ne laser ($633\,{\rm nm}$). The laser beam passes first through a plane polarizer (P) whose axis is parallel to the vertical $y$-axis and a quarter wave plate (QWP) whose optical axis is kept at $\pm45^\circ$ with the vertical axis; depending on the sign of this angle, the resulting light beam is either left circularly polarized (LCP) or right circularly polarized (RCP). Afterward, the beam undergoes diffraction passing through the characteristic ring structure of the POLICRYPS.  Both the diffracted and transmitted beams are then focused onto a CCD camera by a lens. We remark that the part of the beam passing through the  polymer rings does not undergo a change of polarization and propagates as a standard ${\rm TEM}_{00}$ Gaussian beam because the polymer rings are not birefringent (in fact, while some LC molecules can remain trapped in the polymer, they are not aligned); thus, the beam emerging from the POLICRYPS is a superposition of a pure CVB generated by the LC fringes and the unperturbed (Gaussian) beam coming through the polymer rings. This is a drawback in terms of both the quality of the generated CVB and the efficiency of the device. Nevertheless, the polymeric rings in the structure represents a real valuable advantage since they confine and stabilize the LC material, thus preserving the radial alignment even when the device is used with high-power lasers~\cite{Sio1}. 

We illuminated the polshape with a LCP polarized Gaussian beam obtained by adjusting the QWP at $-45^\circ$ with the vertical axis. This generated a diverging CVB, which we could image on a camera by using a lens. Placing the camera at the focal plane of the lens we recorded the beam profile shown in Fig.~\ref{fig:fig4}a. We then proceeded to measure the components of the beams along various linear polarizations by placing the analyzer before the camera at different angles, i.e., $0^\circ$, $45^\circ$, $90^\circ$ and $135^\circ$ (Figs.~\ref{fig:fig4}b-e, the analyzer axis direction is shown by the white arrow in each panel). We finally used this information to infer the local (i.e., as a function of the coordinates over the beam profile) Stokes parameters~\cite{Stokes,Wolf} $I$, $Q$, $U$ and $V$, which completely describe the polarization state of a harmonic electromagnetic field. In fact, the polarization ellipse which characterizes a generic harmonic electric field can be written as
\begin{equation}
I^2=Q^2+U^2+V^2 \; ,
\end{equation}
where the Stokes parameters can be expressed in terms of the field components as
\begin{equation*}
\left\{\begin{array}{ccl}
I & = & E_{0x}^2+E_{0y}^2 \\[6pt]
Q & = & E_{0x}^2-E_{0y}^2 \\[6pt]
U & = & 2E_{0x}E_{0y}\cos(\delta) \\[6pt]
V & = & 2E_{0x}E_{0y}\sin(\delta)
\end{array}\right.
\end{equation*}
where $E_{0x}$ and $E_{0y}$ are the amplitudes of the $x$- and $y$-components of the field, and $\delta$ is their phase difference. These parameters can be represented as a linear combination of the intensity values measured with the analyzer set at orthogonal polarizations, i.e.,
\begin{equation*}
\left\{\begin{array}{ccl}
I & = & I_{0} + I_{90} \\[6pt]
Q & = & I_{0} - I_{90} \\[6pt]
U & = & I_{45}- I_{135} \\[6pt]
V & = & I_{\rm RCP}- I_{\rm LCP} 
\end{array}\right.
\end{equation*}
where $I_0$, $I_{45}$, $I_{90}$ and $I_{135}$ represent the measured intensity at each point for linear polarization of the analyzer at $0^\circ$, $45^\circ$, $90^\circ$ and $135^\circ$, respectively (Figs,~\ref{fig:fig4}b-e), and $I_{\rm RCP}$ and $I_{\rm LCP}$ the intensity obtained using a QWP instead of the analyzer set at $-45^\circ$ and $+45^\circ$, respectively. The polarization angle of the beam is finally given by $\frac{1}{2}\arctan\left(U/Q\right)$ and shown in Fig.~\ref{fig:fig4}f. The resulting intensity and polarization show a cylindrical symmetry permitting us to identify this beam with a hybrid CVB (Fig.~\ref{fig:fig1}c). This hybrid CVB is the result of the interference between the radially polarized CVB diffracted by the POLICRYPS polshape and the undriffracted illuminating Gaussian beam trasmitted through the polshape. Similarly, when we illuminated the POLICRYPS polshape with a RCP Gaussian beam (obtained by adjusting the QWP at $+45^\circ$ with the vertical axis), we generated another hybrid CVB with the same intensity profile, but different polarization (Figs.~\ref{fig:fig4}g-l).

\begin{figure}[!h]
\centering
\includegraphics[width=\linewidth]{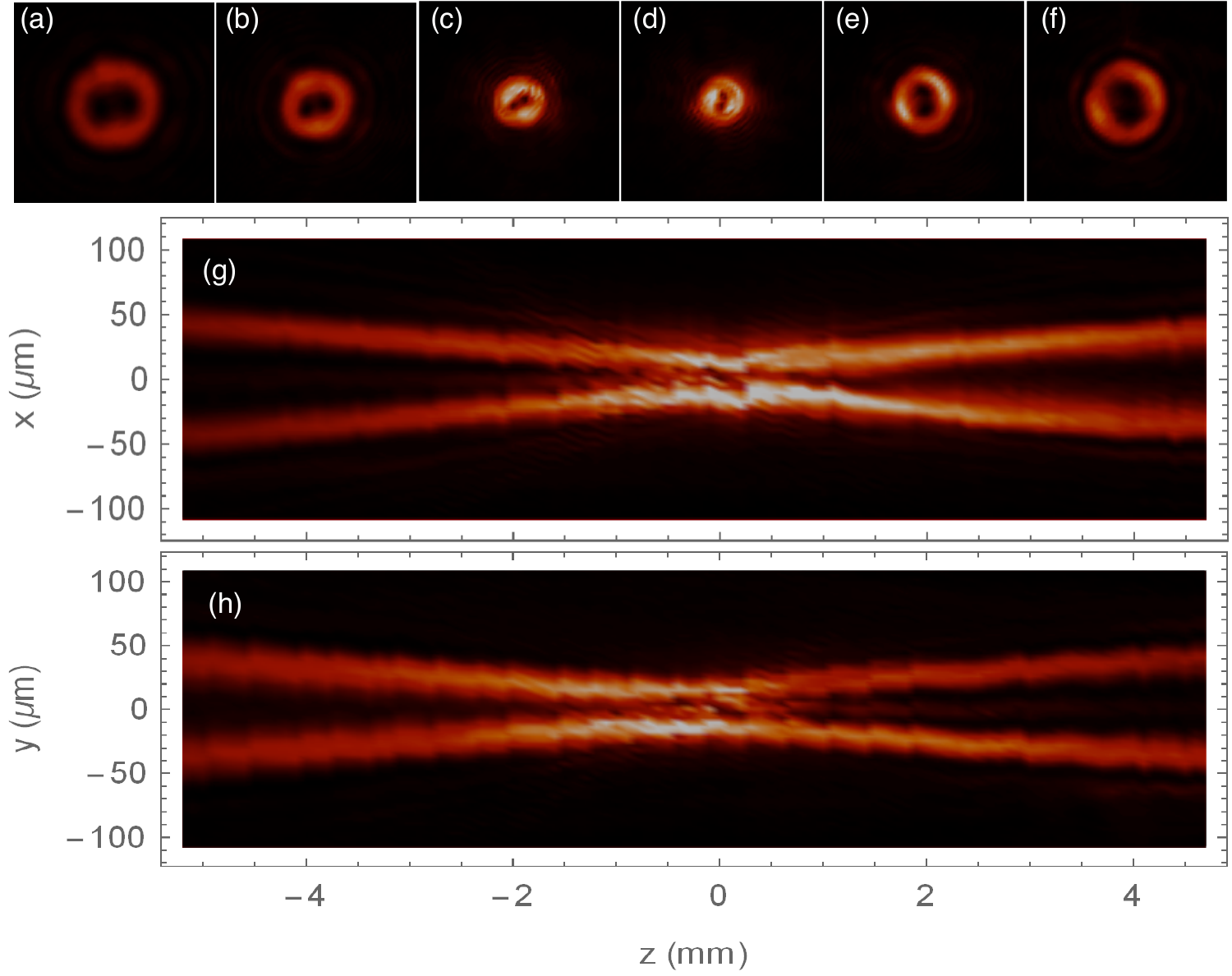}
\caption{(Color online) Propagation of the CVB generated by the POLICRYPS polshape. (a)-(f) transverse beam profiles in the $xy$-plane at $z = -25$, $-15$, $-5$, $5$, $15$ and $25\,{\rm \mu m}$ from the focal plane of the lens (see reference system in Fig.~\ref{fig:fig3}). (g) and (h) beam profiles in the $xz$ ($y=0$) and $yz$ ($x=0$) planes.}
\label{fig:fig5}
\end{figure}

Fig.~\ref{fig:fig5} shows that the transversal intensity profile of the CVB generated by the POLICRYPS polshape is donut-shaped for a quite long path, i.e., at least for $\pm10\,{\rm mm}$ around the focal plane of the lens. This measurement was perfomed by translating the camera (fixed on a single-axis translation stage) along the $z$-direction and acquiring transversal images of the beam profile at $10\,{\rm \mu m}$ steps. 

In conclusion, a cost-effective novel device has been realized that can locally modify the state of polarization of an incoming beam. This has been achieved by fabricating a birefringent (LC) plate with a specific (radial) orientation of the optical axis. This device allows the conversion of a (left or right) circularly polarized beam into a hybrid CVB. This beam can be further transformed into a pure radial or azimuthal CVB using two half-wave plates~\cite{Zhan5,Volpe}. The generated CVB has been experimentally studied by characterizing its  polarization and intensity profiles. We remark that the realized sample represents only one of the possible examples of novel active devices that can locally influence the state of polarization of a light beam. Moreover, considering that the birefringency of the LC material is electrically tunable, application of an external electric field to the device can turn the POLICRYPS polshape into an active device: the polarization shaping ability of the device can be easily controlled by a knob, up to the point to be modified in real-time or completely switched off at will. Experiments of this kind are at present ongoing and the novel possibilities we are actually exploring will be soon available. \\

\begin{acknowledgments}
This work has been possible thanks to two short term scientific missions supported by COST Action IC1208. 
SP was supported by T\"ubitak Grant No. 113Z556.
GV acknowledges funding from Marie Curie Career Integration Grant (MC-CIG) PCIG11GA-2012-321726 and a Distinguished Young Scientist award of the Turkish Academy of Sciences (T\"UBA).
\end{acknowledgments}

\bibliography{bibliography}

\begin{thebibliography}{29}%
\makeatletter
\providecommand \@ifxundefined [1]{%
 \@ifx{#1\undefined}
}%
\providecommand \@ifnum [1]{%
 \ifnum #1\expandafter \@firstoftwo
 \else \expandafter \@secondoftwo
 \fi
}%
\providecommand \@ifx [1]{%
 \ifx #1\expandafter \@firstoftwo
 \else \expandafter \@secondoftwo
 \fi
}%
\providecommand \natexlab [1]{#1}%
\providecommand \enquote  [1]{``#1''}%
\providecommand \bibnamefont  [1]{#1}%
\providecommand \bibfnamefont [1]{#1}%
\providecommand \citenamefont [1]{#1}%
\providecommand \href@noop [0]{\@secondoftwo}%
\providecommand \href [0]{\begingroup \@sanitize@url \@href}%
\providecommand \@href[1]{\@@startlink{#1}\@@href}%
\providecommand \@@href[1]{\endgroup#1\@@endlink}%
\providecommand \@sanitize@url [0]{\catcode `\\12\catcode `\$12\catcode
  `\&12\catcode `\#12\catcode `\^12\catcode `\_12\catcode `\%12\relax}%
\providecommand \@@startlink[1]{}%
\providecommand \@@endlink[0]{}%
\providecommand \url  [0]{\begingroup\@sanitize@url \@url }%
\providecommand \@url [1]{\endgroup\@href {#1}{\urlprefix }}%
\providecommand \urlprefix  [0]{URL }%
\providecommand \Eprint [0]{\href }%
\providecommand \doibase [0]{http://dx.doi.org/}%
\providecommand \selectlanguage [0]{\@gobble}%
\providecommand \bibinfo  [0]{\@secondoftwo}%
\providecommand \bibfield  [0]{\@secondoftwo}%
\providecommand \translation [1]{[#1]}%
\providecommand \BibitemOpen [0]{}%
\providecommand \bibitemStop [0]{}%
\providecommand \bibitemNoStop [0]{.\EOS\space}%
\providecommand \EOS [0]{\spacefactor3000\relax}%
\providecommand \BibitemShut  [1]{\csname bibitem#1\endcsname}%
\let\auto@bib@innerbib\@empty
\bibitem [{\citenamefont {Tabirian}(2013)}]{Tabirian}%
  \BibitemOpen
  \bibfield  {author} {\bibinfo {author} {\bibfnamefont {N.}~\bibnamefont
  {Tabirian}},\ }in\ \href@noop {} {\emph {\bibinfo {booktitle} {$11^{\rm th}$
  Workshop, Novel Optical Materials and Applications}}}\ (\bibinfo {address}
  {Cetraro, Italy},\ \bibinfo {year} {2013})\BibitemShut {NoStop}%
\bibitem [{\citenamefont {Hall}(1996)}]{Hall}%
  \BibitemOpen
  \bibfield  {author} {\bibinfo {author} {\bibfnamefont {D.~G.}\ \bibnamefont
  {Hall}},\ }\href@noop {} {\bibfield  {journal} {\bibinfo  {journal} {Opt.
  Lett.}\ }\textbf {\bibinfo {volume} {21}},\ \bibinfo {pages} {9} (\bibinfo
  {year} {1996})}\BibitemShut {NoStop}%
\bibitem [{\citenamefont {Zhan}(2009)}]{Zhan}%
  \BibitemOpen
  \bibfield  {author} {\bibinfo {author} {\bibfnamefont {Q.}~\bibnamefont
  {Zhan}},\ }\href@noop {} {\bibfield  {journal} {\bibinfo  {journal} {Adv.
  Opt. Photon.}\ }\textbf {\bibinfo {volume} {1}},\ \bibinfo {pages} {1}
  (\bibinfo {year} {2009})}\BibitemShut {NoStop}%
\bibitem [{\citenamefont {Zhan}\ and\ \citenamefont
  {Leger}(2002{\natexlab{a}})}]{Zhan1}%
  \BibitemOpen
  \bibfield  {author} {\bibinfo {author} {\bibfnamefont {Q.}~\bibnamefont
  {Zhan}}\ and\ \bibinfo {author} {\bibfnamefont {J.}~\bibnamefont {Leger}},\
  }\href@noop {} {\bibfield  {journal} {\bibinfo  {journal} {Opt. Express}\
  }\textbf {\bibinfo {volume} {10}},\ \bibinfo {pages} {324} (\bibinfo {year}
  {2002}{\natexlab{a}})}\BibitemShut {NoStop}%
\bibitem [{\citenamefont {Zhan}(2003)}]{Zhan2}%
  \BibitemOpen
  \bibfield  {author} {\bibinfo {author} {\bibfnamefont {Q.}~\bibnamefont
  {Zhan}},\ }\href@noop {} {\bibfield  {journal} {\bibinfo  {journal} {J. Opt.
  A: Pure Appl. Opt.}\ }\textbf {\bibinfo {volume} {5}},\ \bibinfo {pages}
  {229} (\bibinfo {year} {2003})}\BibitemShut {NoStop}%
\bibitem [{\citenamefont {Quabis}\ \emph {et~al.}(2000)\citenamefont {Quabis},
  \citenamefont {Dorn}, \citenamefont {Eberler}, \citenamefont {Gackl},\ and\
  \citenamefont {Leuchs}}]{Quabis}%
  \BibitemOpen
  \bibfield  {author} {\bibinfo {author} {\bibfnamefont {S.}~\bibnamefont
  {Quabis}}, \bibinfo {author} {\bibfnamefont {R.}~\bibnamefont {Dorn}},
  \bibinfo {author} {\bibfnamefont {M.}~\bibnamefont {Eberler}}, \bibinfo
  {author} {\bibfnamefont {O.}~\bibnamefont {Gackl}}, \ and\ \bibinfo {author}
  {\bibfnamefont {G.}~\bibnamefont {Leuchs}},\ }\href@noop {} {\bibfield
  {journal} {\bibinfo  {journal} {Opt. Commun.}\ }\textbf {\bibinfo {volume}
  {179}},\ \bibinfo {pages} {1} (\bibinfo {year} {2000})}\BibitemShut {NoStop}%
\bibitem [{\citenamefont {Dorn}\ \emph {et~al.}(2003)\citenamefont {Dorn},
  \citenamefont {Quabis},\ and\ \citenamefont {Leuchs}}]{Dorn}%
  \BibitemOpen
  \bibfield  {author} {\bibinfo {author} {\bibfnamefont {R.}~\bibnamefont
  {Dorn}}, \bibinfo {author} {\bibfnamefont {S.}~\bibnamefont {Quabis}}, \ and\
  \bibinfo {author} {\bibfnamefont {G.}~\bibnamefont {Leuchs}},\ }\href@noop {}
  {\bibfield  {journal} {\bibinfo  {journal} {Phys. Rev. Lett.}\ }\textbf
  {\bibinfo {volume} {91}},\ \bibinfo {pages} {233901} (\bibinfo {year}
  {2003})}\BibitemShut {NoStop}%
\bibitem [{\citenamefont {Jones}\ \emph {et~al.}(2015)\citenamefont {Jones},
  \citenamefont {Marag\'o},\ and\ \citenamefont {Volpe}}]{Jones}%
  \BibitemOpen
  \bibfield  {author} {\bibinfo {author} {\bibfnamefont {P.~H.}\ \bibnamefont
  {Jones}}, \bibinfo {author} {\bibfnamefont {O.~M.}\ \bibnamefont {Marag\'o}},
  \ and\ \bibinfo {author} {\bibfnamefont {G.}~\bibnamefont {Volpe}},\
  }\href@noop {} {\emph {\bibinfo {title} {Optical Tweezers: Principles and
  Applications}}}\ (\bibinfo  {publisher} {Cambridge University Press},\
  \bibinfo {year} {2015})\BibitemShut {NoStop}%
\bibitem [{\citenamefont {Zhan}(2004{\natexlab{a}})}]{Zhan3}%
  \BibitemOpen
  \bibfield  {author} {\bibinfo {author} {\bibfnamefont {Q.}~\bibnamefont
  {Zhan}},\ }\href@noop {} {\bibfield  {journal} {\bibinfo  {journal} {Opt.
  Express}\ }\textbf {\bibinfo {volume} {12}},\ \bibinfo {pages} {3377}
  (\bibinfo {year} {2004}{\natexlab{a}})}\BibitemShut {NoStop}%
\bibitem [{\citenamefont {Zhan}(2004{\natexlab{b}})}]{Zhan4}%
  \BibitemOpen
  \bibfield  {author} {\bibinfo {author} {\bibfnamefont {Q.}~\bibnamefont
  {Zhan}},\ }\href@noop {} {\bibfield  {journal} {\bibinfo  {journal} {Proc.
  SPIE}\ }\textbf {\bibinfo {volume} {5514}},\ \bibinfo {pages} {275} (\bibinfo
  {year} {2004}{\natexlab{b}})}\BibitemShut {NoStop}%
\bibitem [{\citenamefont {Zhan}\ and\ \citenamefont
  {Leger}(2002{\natexlab{b}})}]{Zhan5}%
  \BibitemOpen
  \bibfield  {author} {\bibinfo {author} {\bibfnamefont {Q.}~\bibnamefont
  {Zhan}}\ and\ \bibinfo {author} {\bibfnamefont {J.~R.}\ \bibnamefont
  {Leger}},\ }\href@noop {} {\bibfield  {journal} {\bibinfo  {journal} {Appl.
  Opt.}\ }\textbf {\bibinfo {volume} {41}},\ \bibinfo {pages} {4630} (\bibinfo
  {year} {2002}{\natexlab{b}})}\BibitemShut {NoStop}%
\bibitem [{\citenamefont {Volpe}\ and\ \citenamefont {Petrov}(2004)}]{Volpe}%
  \BibitemOpen
  \bibfield  {author} {\bibinfo {author} {\bibfnamefont {G.}~\bibnamefont
  {Volpe}}\ and\ \bibinfo {author} {\bibfnamefont {D.}~\bibnamefont {Petrov}},\
  }\href@noop {} {\bibfield  {journal} {\bibinfo  {journal} {Opt. Commun.}\
  }\textbf {\bibinfo {volume} {237}},\ \bibinfo {pages} {89} (\bibinfo {year}
  {2004})}\BibitemShut {NoStop}%
\bibitem [{\citenamefont {Tidwell}\ \emph {et~al.}(1990)\citenamefont
  {Tidwell}, \citenamefont {Ford},\ and\ \citenamefont {Kimura}}]{Tidwell}%
  \BibitemOpen
  \bibfield  {author} {\bibinfo {author} {\bibfnamefont {S.~C.}\ \bibnamefont
  {Tidwell}}, \bibinfo {author} {\bibfnamefont {D.~H.}\ \bibnamefont {Ford}}, \
  and\ \bibinfo {author} {\bibfnamefont {W.~D.}\ \bibnamefont {Kimura}},\
  }\href@noop {} {\bibfield  {journal} {\bibinfo  {journal} {Appl. Opt.}\
  }\textbf {\bibinfo {volume} {29}},\ \bibinfo {pages} {2234} (\bibinfo {year}
  {1990})}\BibitemShut {NoStop}%
\bibitem [{\citenamefont {Erdogan}\ \emph {et~al.}(1992)\citenamefont
  {Erdogan}, \citenamefont {King}, \citenamefont {Wicks}, \citenamefont {Hall},
  \citenamefont {Anderson},\ and\ \citenamefont {Rooks}}]{Erdogan}%
  \BibitemOpen
  \bibfield  {author} {\bibinfo {author} {\bibfnamefont {T.}~\bibnamefont
  {Erdogan}}, \bibinfo {author} {\bibfnamefont {O.}~\bibnamefont {King}},
  \bibinfo {author} {\bibfnamefont {G.~W.}\ \bibnamefont {Wicks}}, \bibinfo
  {author} {\bibfnamefont {D.~G.}\ \bibnamefont {Hall}}, \bibinfo {author}
  {\bibfnamefont {E.~H.}\ \bibnamefont {Anderson}}, \ and\ \bibinfo {author}
  {\bibfnamefont {M.~J.}\ \bibnamefont {Rooks}},\ }\href@noop {} {\bibfield
  {journal} {\bibinfo  {journal} {Appl. Phys. Lett.}\ }\textbf {\bibinfo
  {volume} {60}},\ \bibinfo {pages} {1921} (\bibinfo {year}
  {1992})}\BibitemShut {NoStop}%
\bibitem [{\citenamefont {Stalder}\ and\ \citenamefont
  {Schadt}(1996)}]{Stalder}%
  \BibitemOpen
  \bibfield  {author} {\bibinfo {author} {\bibfnamefont {M.}~\bibnamefont
  {Stalder}}\ and\ \bibinfo {author} {\bibfnamefont {M.}~\bibnamefont
  {Schadt}},\ }\href@noop {} {\bibfield  {journal} {\bibinfo  {journal} {Opt.
  Lett.}\ }\textbf {\bibinfo {volume} {21}},\ \bibinfo {pages} {1948} (\bibinfo
  {year} {1996})}\BibitemShut {NoStop}%
\bibitem [{\citenamefont {Oron}\ \emph {et~al.}(2000)\citenamefont {Oron},
  \citenamefont {Blit}, \citenamefont {Davidson}, \citenamefont {Friesem},
  \citenamefont {Bomzon},\ and\ \citenamefont {Hasman}}]{Oron}%
  \BibitemOpen
  \bibfield  {author} {\bibinfo {author} {\bibfnamefont {R.}~\bibnamefont
  {Oron}}, \bibinfo {author} {\bibfnamefont {S.}~\bibnamefont {Blit}}, \bibinfo
  {author} {\bibfnamefont {N.}~\bibnamefont {Davidson}}, \bibinfo {author}
  {\bibfnamefont {A.~A.}\ \bibnamefont {Friesem}}, \bibinfo {author}
  {\bibfnamefont {Z.}~\bibnamefont {Bomzon}}, \ and\ \bibinfo {author}
  {\bibfnamefont {E.}~\bibnamefont {Hasman}},\ }\href@noop {} {\bibfield
  {journal} {\bibinfo  {journal} {Appl. Phys. Lett.}\ }\textbf {\bibinfo
  {volume} {77}},\ \bibinfo {pages} {3322} (\bibinfo {year}
  {2000})}\BibitemShut {NoStop}%
\bibitem [{\citenamefont {Grosjean}\ \emph {et~al.}(2002)\citenamefont
  {Grosjean}, \citenamefont {Courjon},\ and\ \citenamefont
  {Spajer}}]{Grosjean}%
  \BibitemOpen
  \bibfield  {author} {\bibinfo {author} {\bibfnamefont {T.}~\bibnamefont
  {Grosjean}}, \bibinfo {author} {\bibfnamefont {D.}~\bibnamefont {Courjon}}, \
  and\ \bibinfo {author} {\bibfnamefont {M.}~\bibnamefont {Spajer}},\
  }\href@noop {} {\bibfield  {journal} {\bibinfo  {journal} {Opt. Commun.}\
  }\textbf {\bibinfo {volume} {203}},\ \bibinfo {pages} {1} (\bibinfo {year}
  {2002})}\BibitemShut {NoStop}%
\bibitem [{\citenamefont {Bomzon}\ \emph {et~al.}(2002)\citenamefont {Bomzon},
  \citenamefont {Biener}, \citenamefont {Kleiner},\ and\ \citenamefont
  {Hasman}}]{Bomzon}%
  \BibitemOpen
  \bibfield  {author} {\bibinfo {author} {\bibfnamefont {Z.}~\bibnamefont
  {Bomzon}}, \bibinfo {author} {\bibfnamefont {G.}~\bibnamefont {Biener}},
  \bibinfo {author} {\bibfnamefont {V.}~\bibnamefont {Kleiner}}, \ and\
  \bibinfo {author} {\bibfnamefont {E.}~\bibnamefont {Hasman}},\ }\href@noop {}
  {\bibfield  {journal} {\bibinfo  {journal} {Opt. Lett.}\ }\textbf {\bibinfo
  {volume} {27}},\ \bibinfo {pages} {285} (\bibinfo {year} {2002})}\BibitemShut
  {NoStop}%
\bibitem [{\citenamefont {Kozawa}\ and\ \citenamefont {Sato}(2005)}]{Kozawa}%
  \BibitemOpen
  \bibfield  {author} {\bibinfo {author} {\bibfnamefont {Y.}~\bibnamefont
  {Kozawa}}\ and\ \bibinfo {author} {\bibfnamefont {S.}~\bibnamefont {Sato}},\
  }\href@noop {} {\bibfield  {journal} {\bibinfo  {journal} {Opt. Lett.}\
  }\textbf {\bibinfo {volume} {30}},\ \bibinfo {pages} {3063} (\bibinfo {year}
  {2005})}\BibitemShut {NoStop}%
\bibitem [{\citenamefont {Niziev}\ \emph {et~al.}(2006)\citenamefont {Niziev},
  \citenamefont {Chang},\ and\ \citenamefont {Nesterov}}]{Niziev}%
  \BibitemOpen
  \bibfield  {author} {\bibinfo {author} {\bibfnamefont {V.~G.}\ \bibnamefont
  {Niziev}}, \bibinfo {author} {\bibfnamefont {R.~S.}\ \bibnamefont {Chang}}, \
  and\ \bibinfo {author} {\bibfnamefont {A.~V.}\ \bibnamefont {Nesterov}},\
  }\href@noop {} {\bibfield  {journal} {\bibinfo  {journal} {Appl. Opt.}\
  }\textbf {\bibinfo {volume} {45}},\ \bibinfo {pages} {8393} (\bibinfo {year}
  {2006})}\BibitemShut {NoStop}%
\bibitem [{\citenamefont {Caputo}\ \emph {et~al.}(2004)\citenamefont {Caputo},
  \citenamefont {{De Sio}}, \citenamefont {Veltri}, \citenamefont {Umeton},\
  and\ \citenamefont {Sukhov}}]{Caputo}%
  \BibitemOpen
  \bibfield  {author} {\bibinfo {author} {\bibfnamefont {R.}~\bibnamefont
  {Caputo}}, \bibinfo {author} {\bibfnamefont {L.}~\bibnamefont {{De Sio}}},
  \bibinfo {author} {\bibfnamefont {A.}~\bibnamefont {Veltri}}, \bibinfo
  {author} {\bibfnamefont {C.}~\bibnamefont {Umeton}}, \ and\ \bibinfo {author}
  {\bibfnamefont {A.~V.}\ \bibnamefont {Sukhov}},\ }\href@noop {} {\bibfield
  {journal} {\bibinfo  {journal} {Opt. Lett.}\ }\textbf {\bibinfo {volume}
  {29}},\ \bibinfo {pages} {1261} (\bibinfo {year} {2004})}\BibitemShut
  {NoStop}%
\bibitem [{\citenamefont {Infusino}\ \emph {et~al.}(2012)\citenamefont
  {Infusino}, \citenamefont {Ferraro}, \citenamefont {{De Luca}}, \citenamefont
  {Caputo},\ and\ \citenamefont {Umeton}}]{Infusino}%
  \BibitemOpen
  \bibfield  {author} {\bibinfo {author} {\bibfnamefont {M.}~\bibnamefont
  {Infusino}}, \bibinfo {author} {\bibfnamefont {A.}~\bibnamefont {Ferraro}},
  \bibinfo {author} {\bibfnamefont {A.}~\bibnamefont {{De Luca}}}, \bibinfo
  {author} {\bibfnamefont {R.}~\bibnamefont {Caputo}}, \ and\ \bibinfo {author}
  {\bibfnamefont {C.}~\bibnamefont {Umeton}},\ }\href@noop {} {\bibfield
  {journal} {\bibinfo  {journal} {J. Opt. Soc. Am. B}\ }\textbf {\bibinfo
  {volume} {29}},\ \bibinfo {pages} {3170} (\bibinfo {year}
  {2012})}\BibitemShut {NoStop}%
\bibitem [{\citenamefont {{De Sio}}\ \emph {et~al.}(2013)\citenamefont {{De
  Sio}}, \citenamefont {Veltri}, \citenamefont {Caputo}, \citenamefont {{De
  Luca}}, \citenamefont {Strangi}, \citenamefont {Bartolino},\ and\
  \citenamefont {Umeton}}]{Sio}%
  \BibitemOpen
  \bibfield  {author} {\bibinfo {author} {\bibfnamefont {L.}~\bibnamefont {{De
  Sio}}}, \bibinfo {author} {\bibfnamefont {A.}~\bibnamefont {Veltri}},
  \bibinfo {author} {\bibfnamefont {R.}~\bibnamefont {Caputo}}, \bibinfo
  {author} {\bibfnamefont {A.}~\bibnamefont {{De Luca}}}, \bibinfo {author}
  {\bibfnamefont {G.}~\bibnamefont {Strangi}}, \bibinfo {author} {\bibfnamefont
  {R.}~\bibnamefont {Bartolino}}, \ and\ \bibinfo {author} {\bibfnamefont
  {C.~P.}\ \bibnamefont {Umeton}},\ }\href@noop {} {\bibfield  {journal}
  {\bibinfo  {journal} {Liquid Crystals Rev.}\ }\textbf {\bibinfo {volume}
  {1}},\ \bibinfo {pages} {2} (\bibinfo {year} {2013})}\BibitemShut {NoStop}%
\bibitem [{\citenamefont {Alj}\ \emph {et~al.}(2014)\citenamefont {Alj},
  \citenamefont {Caputo},\ and\ \citenamefont {Umeton}}]{Alj}%
  \BibitemOpen
  \bibfield  {author} {\bibinfo {author} {\bibfnamefont {D.}~\bibnamefont
  {Alj}}, \bibinfo {author} {\bibfnamefont {R.}~\bibnamefont {Caputo}}, \ and\
  \bibinfo {author} {\bibfnamefont {C.}~\bibnamefont {Umeton}},\ }\href@noop {}
  {\bibfield  {journal} {\bibinfo  {journal} {Opt. Lett.}\ }\textbf {\bibinfo
  {volume} {39}},\ \bibinfo {pages} {6201} (\bibinfo {year}
  {2014})}\BibitemShut {NoStop}%
\bibitem [{\citenamefont {Marino}\ \emph {et~al.}(2004)\citenamefont {Marino},
  \citenamefont {Vita}, \citenamefont {Tkachenko}, \citenamefont {Caputo},
  \citenamefont {Umeton}, \citenamefont {Veltri},\ and\ \citenamefont
  {Abbate}}]{Marino}%
  \BibitemOpen
  \bibfield  {author} {\bibinfo {author} {\bibfnamefont {A.}~\bibnamefont
  {Marino}}, \bibinfo {author} {\bibfnamefont {F.}~\bibnamefont {Vita}},
  \bibinfo {author} {\bibfnamefont {V.}~\bibnamefont {Tkachenko}}, \bibinfo
  {author} {\bibfnamefont {R.}~\bibnamefont {Caputo}}, \bibinfo {author}
  {\bibfnamefont {C.}~\bibnamefont {Umeton}}, \bibinfo {author} {\bibfnamefont
  {A.}~\bibnamefont {Veltri}}, \ and\ \bibinfo {author} {\bibfnamefont
  {G.}~\bibnamefont {Abbate}},\ }\href@noop {} {\bibfield  {journal} {\bibinfo
  {journal} {Eur. Phys. J. E}\ }\textbf {\bibinfo {volume} {15}},\ \bibinfo
  {pages} {47} (\bibinfo {year} {2004})}\BibitemShut {NoStop}%
\bibitem [{\citenamefont {Sio}\ \emph {et~al.}(2008)\citenamefont {Sio},
  \citenamefont {Tabiryan}, \citenamefont {Caputo}, \citenamefont {Veltri},\
  and\ \citenamefont {Umeton}}]{Sio1}%
  \BibitemOpen
  \bibfield  {author} {\bibinfo {author} {\bibfnamefont {L.~D.}\ \bibnamefont
  {Sio}}, \bibinfo {author} {\bibfnamefont {N.}~\bibnamefont {Tabiryan}},
  \bibinfo {author} {\bibfnamefont {R.}~\bibnamefont {Caputo}}, \bibinfo
  {author} {\bibfnamefont {A.}~\bibnamefont {Veltri}}, \ and\ \bibinfo {author}
  {\bibfnamefont {C.}~\bibnamefont {Umeton}},\ }\href@noop {} {\bibfield
  {journal} {\bibinfo  {journal} {Opt. Express}\ }\textbf {\bibinfo {volume}
  {16}},\ \bibinfo {pages} {7619} (\bibinfo {year} {2008})}\BibitemShut
  {NoStop}%
\bibitem [{\citenamefont {Caputo}\ \emph {et~al.}(2010)\citenamefont {Caputo},
  \citenamefont {Trebisacce}, \citenamefont {{De Sio}},\ and\ \citenamefont
  {{Umeton}}}]{Caputo1}%
  \BibitemOpen
  \bibfield  {author} {\bibinfo {author} {\bibfnamefont {R.}~\bibnamefont
  {Caputo}}, \bibinfo {author} {\bibfnamefont {I.}~\bibnamefont {Trebisacce}},
  \bibinfo {author} {\bibfnamefont {L.}~\bibnamefont {{De Sio}}}, \ and\
  \bibinfo {author} {\bibfnamefont {C.}~\bibnamefont {{Umeton}}},\ }\href@noop
  {} {\bibfield  {journal} {\bibinfo  {journal} {Opt. Express}\ }\textbf
  {\bibinfo {volume} {18}},\ \bibinfo {pages} {5776} (\bibinfo {year}
  {2010})}\BibitemShut {NoStop}%
\bibitem [{\citenamefont {Stokes}(1852)}]{Stokes}%
  \BibitemOpen
  \bibfield  {author} {\bibinfo {author} {\bibfnamefont {G.~G.}\ \bibnamefont
  {Stokes}},\ }\href@noop {} {\bibfield  {journal} {\bibinfo  {journal} {Trans.
  Cambridge Phil. Soc.}\ }\textbf {\bibinfo {volume} {9}},\ \bibinfo {pages}
  {399} (\bibinfo {year} {1852})}\BibitemShut {NoStop}%
\bibitem [{\citenamefont {Born}\ and\ \citenamefont {Wolf}(1988)}]{Wolf}%
  \BibitemOpen
  \bibfield  {author} {\bibinfo {author} {\bibfnamefont {M.}~\bibnamefont
  {Born}}\ and\ \bibinfo {author} {\bibfnamefont {E.}~\bibnamefont {Wolf}},\
  }\href@noop {} {\emph {\bibinfo {title} {Principles of Optics}}}\ (\bibinfo
  {publisher} {Cambridge University Press},\ \bibinfo {year}
  {1988})\BibitemShut {NoStop}%
\end{thebibliography}%

\end{document}